\documentclass{PoS}
\let\ifpdf\relax
\pdfoutput=1
\usepackage{amsmath}
\usepackage{graphicx}
\usepackage{setspace}

\newcommand{\Dlr}{{D^{\hspace{-0.8em}%
      \raisebox{0.8ex}{$\scriptstyle\leftrightarrow$}}}{}}
\newcommand{\Dl}{{D^{\hspace{-0.8em}%
      \raisebox{0.8ex}{$\scriptstyle\leftarrow$}}}{}}
\newcommand{\Dr}{{D^{\hspace{-0.8em}%
      \raisebox{0.8ex}{$\scriptstyle\rightarrow$}}}{}}
      
\newcommand{\Fig}[1]{\mbox{Fig.\,\ref{#1}}}

\newcommand{\preprint}{
  \begin{picture}(0,0)
    \put(0,90){{\rm\normalsize DESY 12-038, Edinburgh 2012/03, LTH 941}}
  \end{picture}}

\title{\preprint
  First moments of the nucleon generalized parton distributions
  from lattice QCD} 

\makeatletter \ShortTitle{\@title} \makeatother   

\author{\speaker{A.~Sternbeck}$^1$\thanks{Supported by the
          FP7-People Programme of the European Commission.},
          \,M.~G\"ockeler$^1$,
          Ph.~H\"agler$^2$,
          R.~Horsley$^3$,
          Y.~Nakamura$^4$,
          A.~Nobile$^5$,
          D.~Pleiter$^{1,5}$,
          P.E.L.~Rakow$^6$,
          A.~Sch\"afer$^1$,
          G.~Schierholz$^7$,
          J.~Zanotti$^8$ \\
         ${}^1$ Institut f\"ur Theoretische Physik, Universit\"at
                Regensburg, 93040 Regensburg, Germany\\
         ${}^2$ Institut f\"ur Kernphysik, Johannes Gutenberg-Universit\"at
                Mainz, 55128 Mainz, Germany\\
         ${}^3$ School of Physics and Astronomy, University of Edinburgh,
                Edinburgh EH9 3JZ, UK \\
         ${}^4$ RIKEN Advanced Institute for Computational Science, Kobe, 
                Hyogo 650-0047, Japan \\
         ${}^5$ JSC, J\"ulich Research Centre, 52425 J\"ulich, Germany\\
         ${}^6$ Theoretical Physics Division, Department of Mathematical
                Sciences, University of Liverpool, Liverpool L69 3BX, UK\\
         ${}^7$ Deutsches Elektronen-Synchrotron DESY, 22603 Hamburg, Germany\\
         ${}^8$ School of Chemistry and Physics, University of Adelaide, SA
          5005, Australia\\
         E-mail: \email{andre.sternbeck@ur.de}}

\abstract{%
We report on our lattice calculations of the nucleon's generalized parton
distributions (GPDs), concentrating on their first moments for the case of
$N_f=2$. Due to recent progress on the numerical side we are able to present
results for the generalized form factors at pion masses as low as 260 MeV. We
perform a fit to one-loop covariant baryon chiral perturbation theory with
encouraging results.}

\FullConference{The XXIX International Symposium on Lattice Field
  Theory - Lattice 2011\\ 
  July 10-16, 2011\\
  Squaw Valley, Lake Tahoe, California}

\begin{document}

\section{Introduction}

The study of the internal structure of hadrons still presents an exciting
challenge. Among the different types of studies, the computation of Generalized
Parton Distributions (GPDs) is especially challenging, but also attractive,
because of their potential for hadron physics.

GPDs were introduced in the late 90s. For a given hadron, they provide 
detailed information on the partonic structure with respect to spatial,
momentum and spin degrees of freedom. GPDs combine the
information of the traditional form factors and parton distribution functions
(containing them as limiting cases) into a single set of functions and hence
contain information also on the correlation between the momentum, spin
and spatial degrees of freedom. For the nucleon, one hopes GPDs will provide one
day a three-dimensional spatial picture, a better understanding of its spin
structure and a value for the quark orbital angular momentum
\cite{Belitsky:2005qn}.

Beside the renormalization scale\footnote{For simplicity, we drop the explicit
reference to the renormalization\ scale $\mu$ in what follows. It is always
implicitly understood. Our lattice data below has been nonperturbatively
renormalized \cite{Gockeler:2010yr} 
and is for the $\overline{\mathsf{MS}}$ scheme at
$\mu=2\,\text{GeV}$.}, GPDs depend on three
kinematic variables: the longitudinal parton momentum fraction $x$, the
skewness parameter $\xi$ and the virtuality $t$. 
The quark structure of a nucleon, for example, is governed by eight GPDs.
Among these, the unpolarized GPDs $H$ and $E$ parametrize the off-diagonal
matrix element\vspace{-0.5ex}
\begin{equation}
  \label{eq:GPDs}
   \langle N(P')|\mathcal{O}_V^\mu(x)|N(P)\rangle =
  \overline U(P')\left\{\gamma^\mu H(x,\xi,t)+\frac{ i\sigma^{\mu\nu}\Delta_\nu}
{2m_N}     
   E(x,\xi,t)\right\}U(P)\;+\textrm{higher twist}\,.
\end{equation}
Here $P$ and $P'$ denote the incoming and outgoing nucleon
momenta (and so $\Delta=P'-P$, $\overline{P}=(P'+P)/2$, $t=\Delta^2$ and
$\xi=-n\cdot\Delta/2$)
and $\mathcal{O}^\mu_V(x)$ is the light-cone bilocal operator\vspace{-0.5ex}
\begin{equation}
\label{eq:bilocal_op}
 \mathcal{O}^\mu_V(x) = \int^\infty_{-\infty}\frac{d\lambda}{2\pi} e^{i\lambda
x}
   \; \bar{q}\left(-\frac{\lambda}{2}n\right) \gamma^\mu \mathcal{P}
e^{-ig\int^{-\lambda/2}_{\lambda/2}d\alpha\, n  A(\alpha n)}
  q \left(\frac{\lambda}{2}n\right)\,,
\end{equation} 
which often arises in hard scattering processes (see, e.g., \cite{Ji:1998pc}).
$n$ (with $\overline{P}\cdot n=1$) denotes a light cone vector in 
Eq.\eqref{eq:bilocal_op} and $\mathcal{P}$
the correct path-ordering of the gluon fields $A$. 
The polarized nucleon GPDs, $\tilde{H}$ and $\tilde{E}$, are defined in a
similar manner, replacing $\gamma^\mu$ in Eq.\eqref{eq:bilocal_op} by
$\gamma^\mu\gamma_5$.\vspace{-1ex}

\section{GPDs and the lattice}

GPDs can be accessed experimentally, for instance, via deeply virtual
Compton scattering. The analysis, however, is rather demanding and requires
also a partial modeling of the combined $x$-, $\xi$- and $t$-dependence.
Cross-checks to other methods are thus inevitable. 

A promising method is given by lattice QCD computations. Although a direct
determination of GPDs on the lattice is not possible, their (Mellin)
moments\vspace{-1ex}
\begin{equation}
 \int_{-1}^1 dx\, x^{n-1} H(x,\xi,t),\quad \int_{-1}^1 dx\, x^{n-1} E(x,\xi,t),
\quad\ldots 
\end{equation}
are accessible. For a nucleon, for example, these moments can be calculated
via matrix elements $\langle N(P')|O|N(P)\rangle$ of local
operators $O$. For $H$ and $E$ these operators read
\begin{equation}
  O_{V}^{\mu\nu_1\cdot\nu_{n-1}}(z)
   = \overline{q}(z) \, \gamma^{\{\mu}
      i \Dlr^{\nu_1} \cdots i\Dlr^{\nu_{n-1}\}} q(z) -\text{traces}
  \label{eq:O}
\end{equation}
where $q$ refers to a quark field, $\Dlr\equiv\Dr-\Dl$ to the covariant
derivative and $\{\cdots\}$ to a symmetrization of the Lorentz indices. For a
definition and further details on the operators needed for the remaining nucleon
GPDs, the reader may refer to \cite{Hagler:2009ni}.

Admittedly, the computation of such matrix elements is quite demanding
already for $n\ge2$, and we are not yet in the stage to provide precision 
results close to the physical point. Nonetheless, such calculations have become
more and more feasible in recent years, and hence have attracted interest from
within the lattice community
\cite{Gockeler:2003jfa,Hagler:2003jd,Hagler:2003is,Hagler:2007xi,
Brommel:2007xd,Bratt:2010jn,Alexandrou:2011nr}.

In what follows, we will restrict ourselves to the two nucleon GPDs $H$ and $E$.
Their moments are polynomials in $\xi$,\vspace{-1.5ex}
\begin{equation}
  \label{eq:formfactors}
  \int_{-1}^1 dx\, x^{n-1} \left[\begin{array}{c} 
    H(x,\xi,t) \\ E(x,\xi,t) \end{array}\right]
= \sum_{k=0}^{[(n-1)/2]}
(2\xi)^{2k} \left[\begin{array}{c}
 A_{n,2k}(t) \\ B_{n,2k}(t)\end{array}\right] 
\pm \delta_{n,\rm even}(2\xi)^{n} C_{n}(t)\,.
\end{equation}
The expansion coefficients $A$, $B$ and $C$ are real functions of the
momentum transfer $t$ (and the renormalization scale $\mu$) and are
known as the \emph{Generalized Form Factors} (GFFs) of the nucleon. 
In this notation, for instance, $A_{10}$ and $B_{10}$ correspond to the
electromagnetic form factors \cite{Collins:2011mk}, and $A_{20}$,
$B_{20}$ and $C_{2}$ parametrize the matrix elements of the
energy-momentum tensor $O_V^{\mu\nu}$\vspace{-0.5ex}
\begin{equation}
  \label{eq:Tmunu}
  \langle N(P')|O^{\mu\nu}_V|N(P)\rangle = \overline{U}(P')\bigg\{\gamma^{\{\mu}
\overline P^{\nu\}}
    A_{20}(t) - \frac{ i\Delta_\rho\sigma^{\rho\{\mu}} {2m_N} \overline 
P^{\nu\}} B_{20}(t)
    +\frac{\Delta^{\{\mu}\Delta^{\nu\}}}{m_N} C_{2}(t) \bigg\}U(P)\;. 
\end{equation}

Below we present results for $A_{20}$, $B_{20}$ and $C_{2}$. They can be
extracted from ratios of two- and three-point correlation functions\vspace{-2ex}
\begin{equation}
  R(t,\tau,p',p) =  
 \frac{C_3(t,\tau,p',p)}{C_2(t,p')} \times
  \left[ \frac{C_2(\tau,p') C_2(t,p') C_2(t-\tau,p) }
 {C_2(\tau,p) C_2(t,p) C_2(t-\tau,p')} 
 \right]^{1/2}\,,
\end{equation}
which are proportional to $\langle N(P')|O^{\mu\nu}_V|N(P)\rangle$ and
constant in the limit $0 \ll \tau \ll t \lesssim T/2$ ($T$ is the temporal
lattice extension). $C_2(t,p)$ is the nucleon two-point
function with a source at time $0$ and sink at time $t$, and $C_3(t,\tau,p',p)$
is the three-point function with an operator insertion at time $\tau$. The
latter we calculate employing the sequential source technique.
\vspace{-1.5ex}

\begin{floatingfigure}[r]
  \begin{minipage}[b]{0.45\textwidth}  
\includegraphics[width=\textwidth]{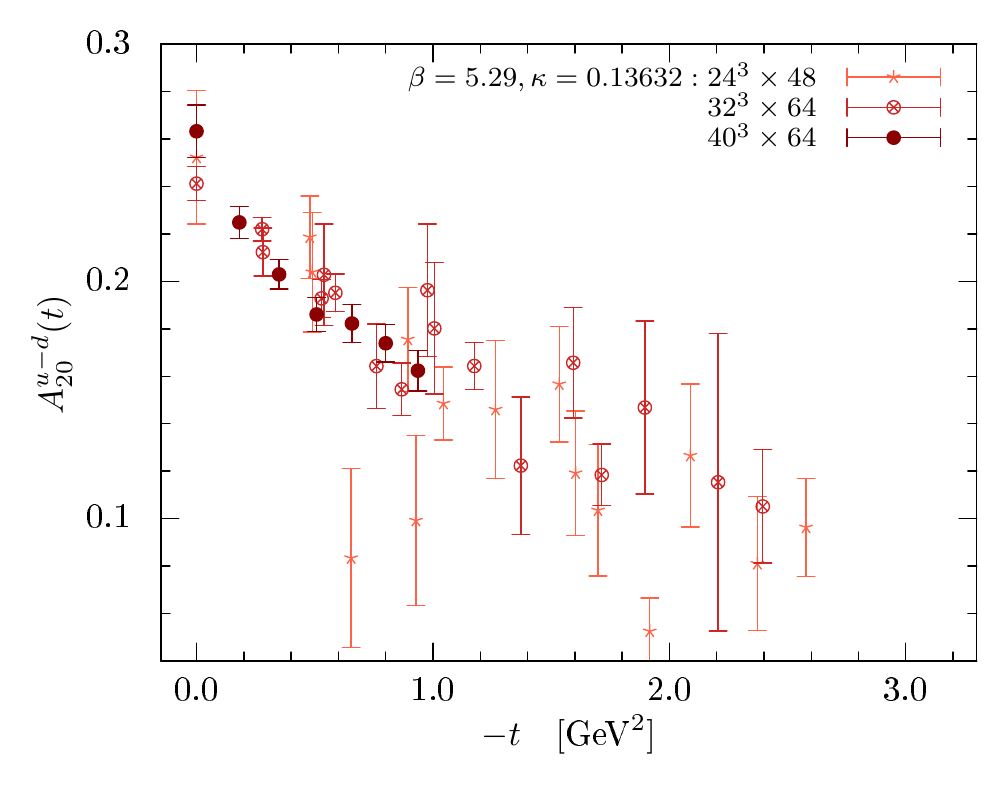}
 \vspace{-6ex}
   \caption{$A_{20}(t)$ for the isovector case for $\beta=5.29$ and
    $\kappa=0.13632$ and for different volumes.} 
   \label{fig:A20_v_5p29_finV}   
    \end{minipage}
 \end{floatingfigure} 
\section{Results}

\begin{figure*}
  \centering
  \mbox{%
    \includegraphics[width=0.5\textwidth]{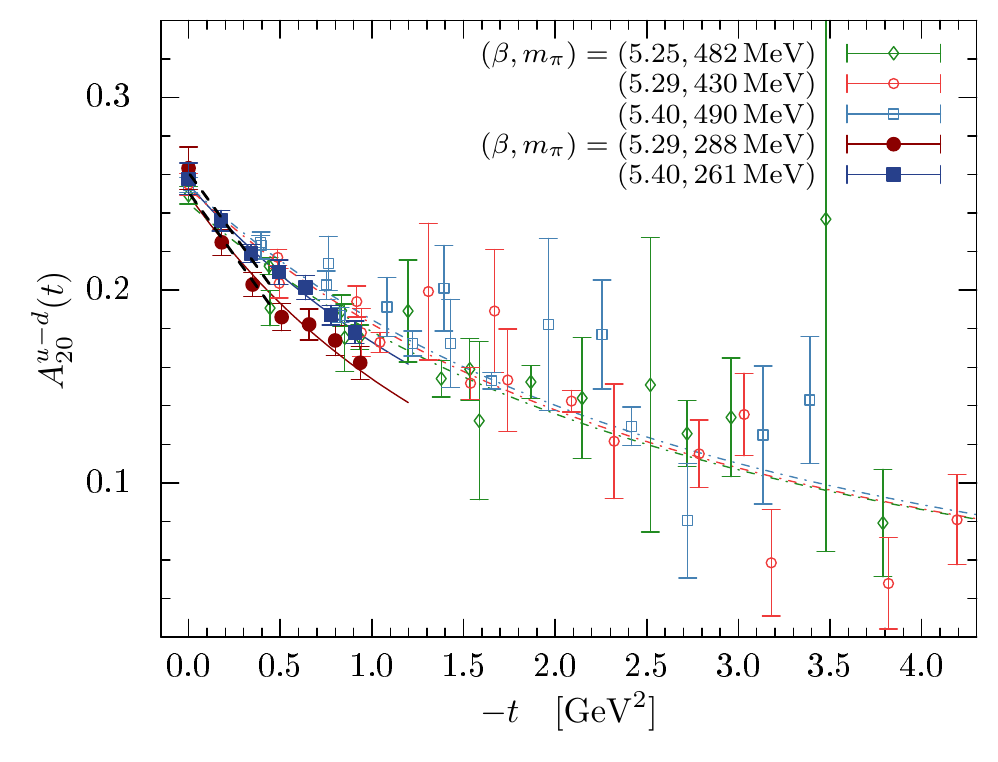}%
    \includegraphics[width=0.5\textwidth]{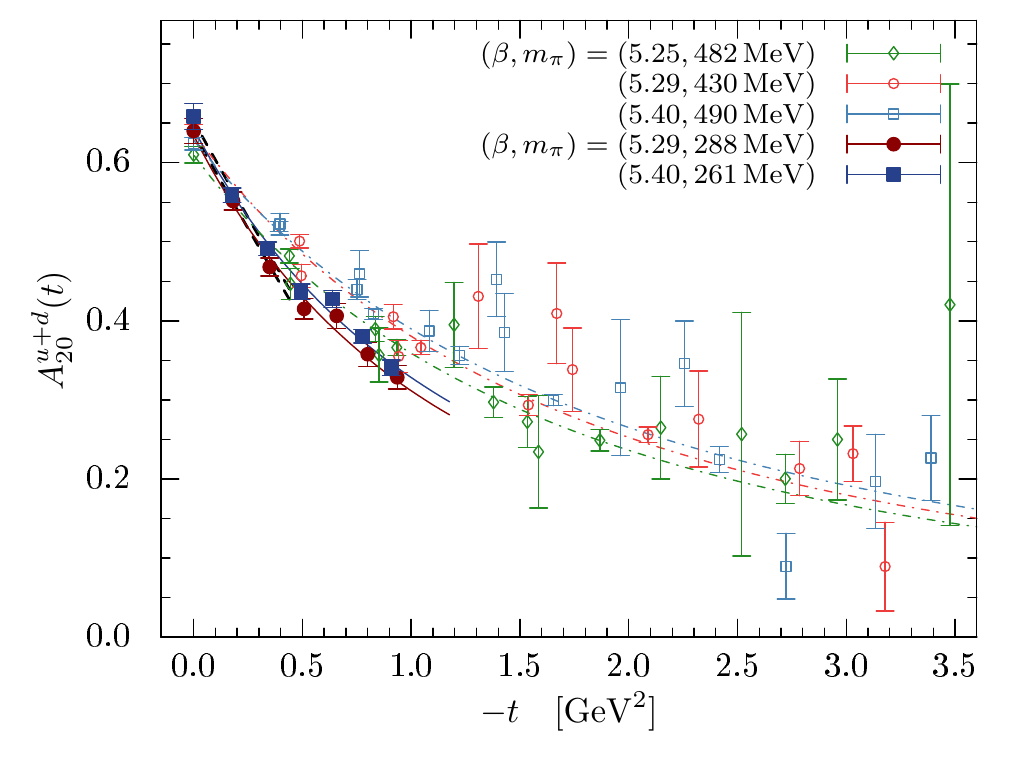}}
  \mbox{%
    \includegraphics[width=0.5\textwidth]{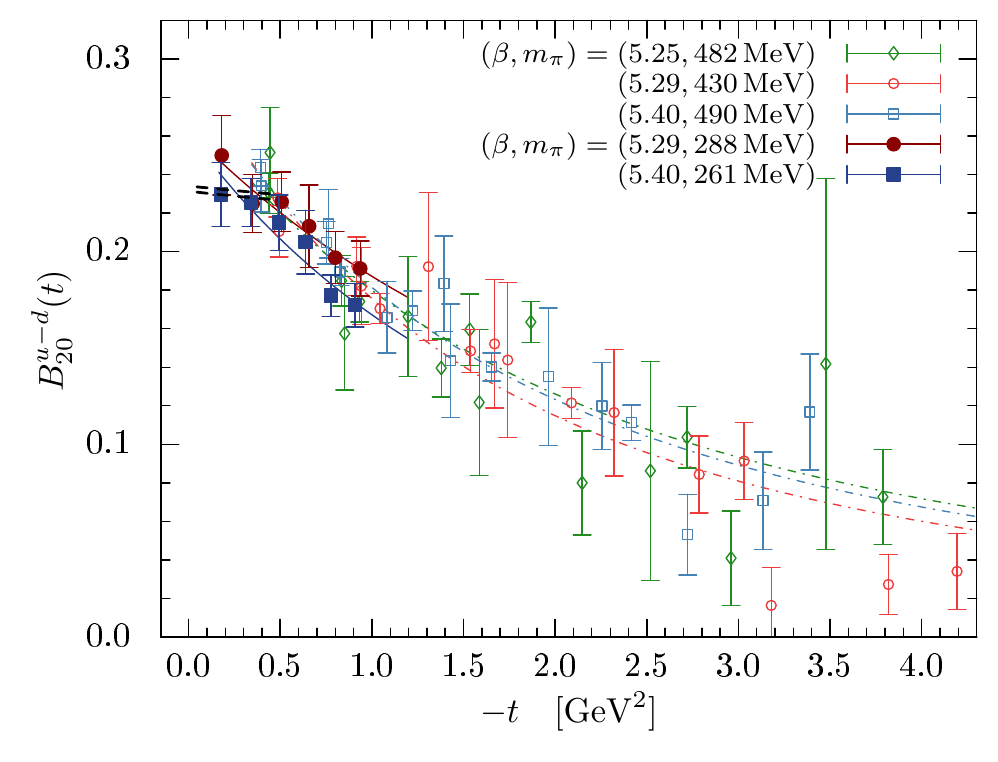}%
    \includegraphics[width=0.5\textwidth]{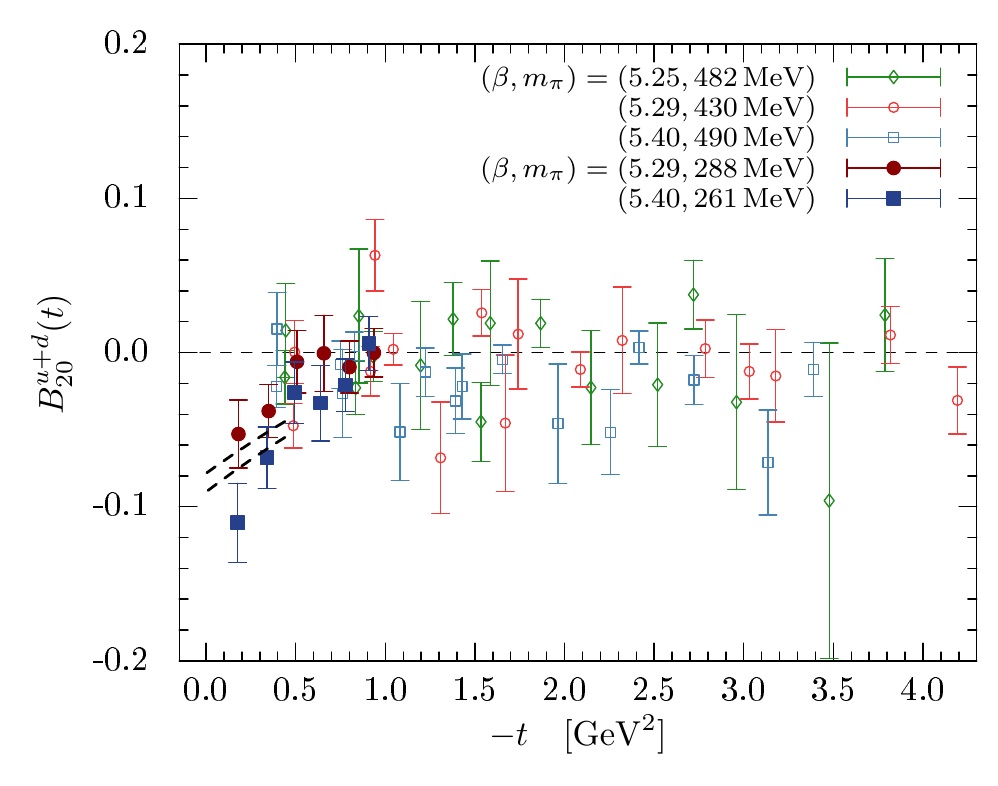}}
  \mbox{%
    \includegraphics[width=0.5\textwidth]{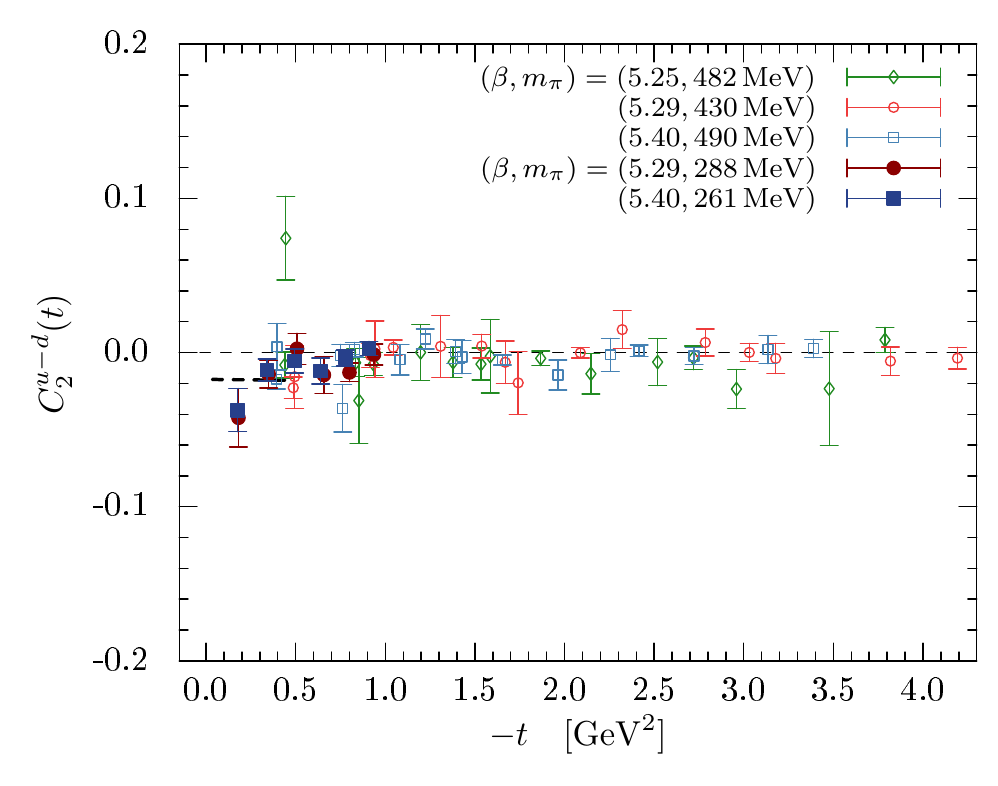}%
    \includegraphics[width=0.5\textwidth]{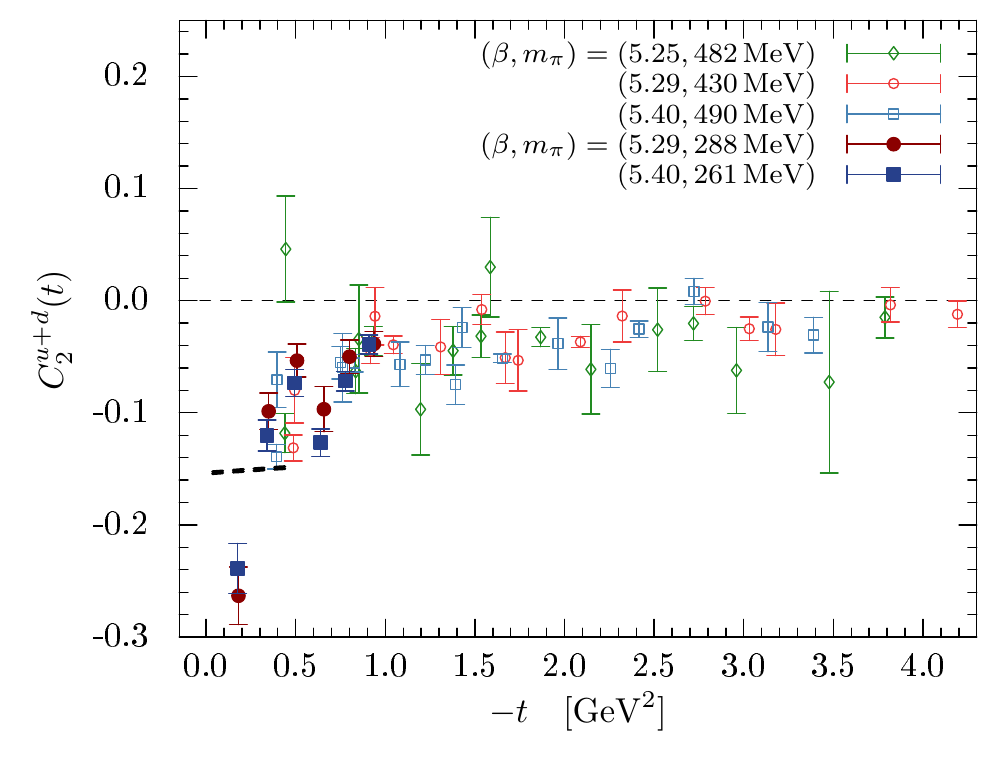}}
  \caption{The generalized form factors $A_{20}$, $B_{20}$ and $C_{2}$
   (from top to bottom) vs.\ momentum transfer $-t$; left for the
   isovector channel, right for the isosinglet channel. The data is for three
   lattice spacings and two groups of approximately equal pion masses. If
   applicable, solid (dashed-dotted) lines represent dipole fits to the data.
   Dashed lines at low $t$ result from a \underline{simultaneous} fit of the
   low-pion-mass data (full symbols) to covariant chiral perturbation theory
   (see text and also \protect\Fig{fig:A20_vs_t_zoom_in} for more details).} 
  \label{fig:GFF_vs}
\end{figure*}

\vspace{-1ex}
Our data for the GFFs is for gauge configurations thermalized using the standard
Wilson gauge action and two flavors of clover-improved Wilson fermions. The
gauge couplings are $\beta=5.25$, 5.29 and 5.40; and the $\kappa$ values are
such that pion masses from $1\,\text{GeV}$ down to $260\,\text{MeV}$ are
simulated, where we primarily work with the data in the mass range
$260\,\text{MeV}\le m_\pi \le 490\,\text{MeV}$. The scale is fixed through
setting $r_0=0.5\,\text{fm}$. This is about the value we obtain from chiral
extrapolations of our nucleon mass data \cite{Bali:2012inpreparation} for the
same set of configurations. The lattice sizes are $24^3\times48$,
$32^3\times64$, $40^3\times64$ and $48^3\times64$. In particular the latter two
provide us with a good signal-to-noise ratio. See, for example,
\Fig{fig:A20_v_5p29_finV}, where data for $A_{20}(t)$ in the isovector
channel is shown for the lattice sizes $24^3\times48$, $32^3\times64$ and
$40^3\times64$ at $\beta=5.29$ and $\kappa=0.13632$ ($m_\pi=287$~MeV). The
number of measurements
is 2755, 3013 and 1478, respectively. The advantage of volume averaging is
clearly evident as with about half the statistics, the data
for the $40^3\times64$ lattice comes with much less statistical noise than that
for the $24^3\times48$ lattice. Moreover, \Fig{fig:A20_v_5p29_finV}
indicates that finite size effects are small, at least at our level of
precision.

\begin{floatingfigure}
 \begin{minipage}[t]{0.47\textwidth}
\includegraphics[height=7.5cm]{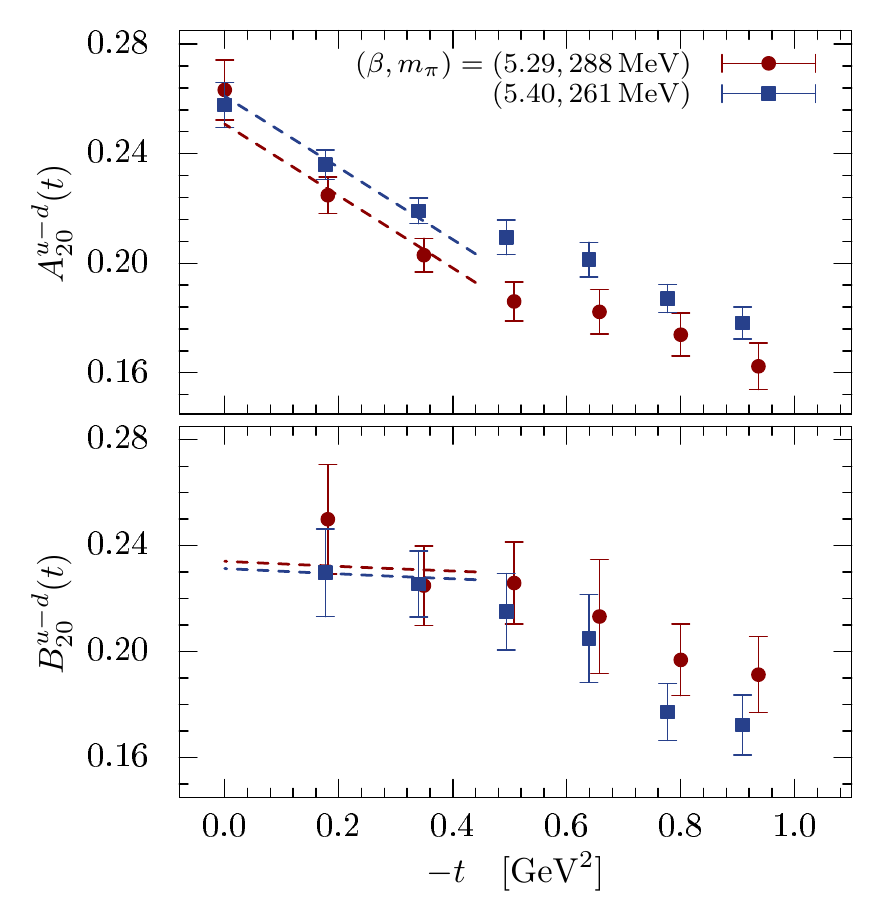}
 \vspace{-3.5cm}
  \caption{$A^{u-d}_{20}$ (top) and $B^{u-d}_{20}$ (bottom) vs.\
$-t$
    at the pion masses 261 and 288\,MeV. Dashed lines result from a
    simultaneous fit of the data (including that for $C^{u-d}_{2}$)
    to expectations from BChPT \cite{Dorati:2007bk}.} 
  \label{fig:A20_vs_t_zoom_in}
  \end{minipage}
\end{floatingfigure}
A selection of all of our GFF data is shown as a function of the momentum
transfer $-t$ in \Fig{fig:GFF_vs}. There, the panels from top to bottom display
the respective data for $A_{20}(t)$, $B_{20}(t)$ and $C_{2}(t)$. Left panels are
for the isovector case, right panels for the isoscalar (without disconnected
contributions). For simplicity, only data for five ensembles is shown, which
fall into two groups of approximately equal pion mass. For the larger pion mass
(430--490\,MeV), we have results for three lattice spacings ($a=0.06$, 0.07 and
0.08\,fm), for the smaller one (260--287\,MeV) we can show data for two sets
($a=0.06,0.07$\,fm). For this (admittedly small) range of lattice spacings we
observe, however, no systematic dependence on $a$. Apparently, there is a slight
vertical shift in the data for $A_{20}(t<0)$ [and in the opposite direction for
$B_{20}(t<0)$] for the lighter sets, but we do not see these shifts for the
heavier sets (including those not shown), at least with the available
precision. It will be interesting to see how well our forthcoming results at
$\beta=5.25$, $\kappa=0.13620$ (i.e., $a=0.084$\,fm, $260$\,MeV pion mass) fit
to these findings.

Similarly, we observe a trend for $A_{20}(t)$ if the pion mass is changed: The
low-$t$ dependence of $A_{20}(t)$ gains slope if $m_{\pi}$ is reduced from
430--490\,MeV to 260--287\,MeV. This effect, however, is small and we see no
such effect in the data from 500\,MeV to 1\,GeV pion mass. It thus remains to be
seen if this effect at lower $m_{\pi}$ stays or disappears with higher
statistics. As above, a further check should become possible as soon as our
results at $\beta=5.25$ and $\kappa=0.13620$ are available.

We can confirm though the (notorious) weak $m_{\pi}$ dependence
of $A^{u-d}_{20}$ at $t=0$, i.e., of $\langle x\rangle^{u-d}$ (see upper
left panel in \Fig{fig:GFF_vs}). From phenomenology one expects $\langle
x\rangle^{u-d}\approx0.16$ at the physical point. So far, however, all available
(world) lattice data for $\langle x\rangle^{u-d}$ for pion masses above 200 MeV
gives values for $\langle x\rangle^{u-d}$ well above 0.16, and moreover,
almost no signal for a downwards trend towards the physical point is seen 
(see, e.g., \cite{Hagler:2009ni} and references therein). From
baryon chiral perturbation theory (BChPT), for example, such a trend is
expected, but it has not yet been demonstrated (convincingly) on the lattice.

It is however interesting that our data for $|t|<0.4\text{GeV}^2$ indicates an
almost linear $t$-dependence for $A^{u-d}_{20}(t)$ and a flattening of the slope
for $B^{u-d}_{20}(t)$. This would be consistent with expectations from covariant
BChPT at leading-one-loop order \cite{Dorati:2007bk}.

In \Fig{fig:GFF_vs}, and in particular in \Fig{fig:A20_vs_t_zoom_in}, we show
a first attempt of fitting our data to the BChPT expressions for $A_{20}$,
$B_{20}$ and $C_2$ as worked out in \cite{Dorati:2007bk} (see the dashed lines
at lower $t$). Note that such a fit has to be a combined fit to the data for
all three GFFs simultaneously, because the parameter $a_{20}$ enters all of
them. The dashed lines in Figs.~\ref{fig:GFF_vs} and \ref{fig:A20_vs_t_zoom_in}
refer to such a fit which incorporates only the lighter data sets (full symbols)
and points for $|t|<0.44\,\textrm{GeV}^2$. Five parameters ($a_{20}$,
$b_{20}$ $c_{20}$, $c_{8}^r$ and $c_{12}$) were left free, while
the phenomenological value $\langle \Delta
x\rangle_{u-d}^{\mathrm{phen}}=0.21$ was used to constrain the coupling $\Delta
a^v_{20}$. The $m_\pi$-dependence of the nucleon mass (entering the BChPT
expression for $B_{20}$ and $C_2$) and the parameter $M_0$ were taken from
our nucleon mass fits \cite{Bali:2012inpreparation}.

It turns out that for the isovector case the fit quality is quite good:
A reduced $\chi^2$-value of about one is reached and the low-$t$ dependence of
$A_{20}$ and $B_{20}$ is roughly reproduced; actually, also for $C_2$, as only
the data point at the smallest $|t|$ falls somewhat below the fitting curve. For
the isoscalar case, the same fit works less satisfactorily. The reason might be
that BChPT does not work at the pion masses under consideration, or that
disconnected contributions are not negligible. The latter are certainly worth
to be calculated. 

\begin{floatingfigure}
  \begin{minipage}[t]{0.45\textwidth}\vspace{-0.4cm}
   \includegraphics[width=1.05\textwidth]{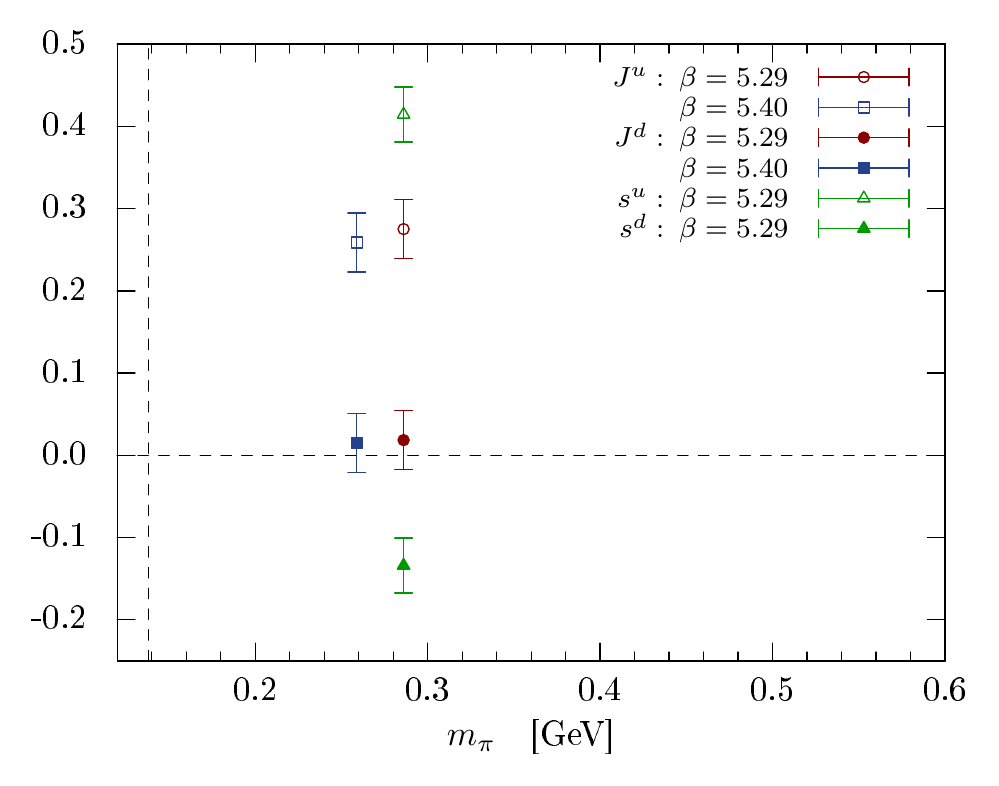}\vspace{-0.5cm}
   \caption{The total quark angular momentum $J$ and
     spin $s^q$ vs.\ pion mass.}
   \label{fig:j_vs_mpi}
  \end{minipage}  
\end{floatingfigure}
Even though disconnected contributions are still missing, it is interesting
to look at the total quark angular momentum
\begin{equation}
 J^{u,d} = \frac{1}{2}\left[A_{20}^{u,d}(0) + B^{u,d}_{20}(0)\right]\,,
\end{equation}
to check if it is in the ballpark of expected values. In
\Fig{fig:j_vs_mpi} we
show this data for our lighter sets, that is, for pion masses 261 and
287\,MeV. Note that for $A^{u\pm d}_{20}$ we have data directly at $t=0$, but
not for $B^{u\pm d}_{20}(0)$. However, looking at \Fig{fig:GFF_vs} one
easily sees that the main contribution to $J^{u+d}$ comes from $A^{u+d}_{20}(0)$
and the $t$-dependence of $B^{u\pm d}_{20}$ is comparably weak. We therefore
approximate $B^{u\pm d}_{20}(0)$ by our data for the smallest $\vert t\vert$.
This should be perfectly fine for our purposes, given all the other
uncertainties and the lack of disconnected contributions.
Note that in \Fig{fig:j_vs_mpi} we have also included data for the quark spin
\vspace{-1ex}
\begin{equation}
 \label{eq:At10}
 s^q = \frac{1}{2}\int^1_{-1} dx\, \tilde{H}(x,\xi,0) = \frac{1}{2}
\tilde{A}^q_{10}(t=0)\,,
\end{equation} 
which we obtain from data\footnote{Unfortunately, there is no data for
$\tilde{A}^{u\pm d}_{10}$ at $\beta=5.4$ for these small pion masses,
but it will become available soon.} for the axial nucleon GFFs
$\tilde{A}^{u-d}_{10}$ and $\tilde{A}^{u+d}_{10}$. 

If we compare our data in \Fig{fig:j_vs_mpi} with that of the LHPC
collaboration \cite{Syritsyn:2011vk}, we find good agreement (albeit their
data is for $N_f=2+1$). We also see the same ordering for the total and
orbital ($L^q=J^q-s^q$) angular momentum and the quark spin:
\begin{displaymath}
 |J^d|\ll |J^u|, \quad |J^d| \ll |s^d|, \quad |L^{u+d}|\ll
|L^u|,|L^d|\,.
\end{displaymath}
It will be interesting to see how this figure changes when data at
smaller pion masses becomes available and/or disconnected contributions are
included.

\section{Conclusions}

We have presented an update on our efforts to calculate the
generalized form factors for the nucleon. We have restricted ourselves here to
the case of $N_f=2$ and reported only on results for the GFFs of the
energy-momentum tensor ($n=2$). Due to recent progress on the numerical side we
are able to provide data for these GFFs for pion masses down to 260\,MeV. In
particular our lighter sets provide an improvement of the available data for
these form factors: Large lattice volumes have allowed us to obtain a very good
signal-to-noise ratio, and at low $|t|$ our data starts to fulfill 
expectations from one-loop BChPT. When comparing our data to that of the LHPC
collaboration presented at this conference \cite{Syritsyn:2011vk} we see a small
vertical offset for the GFF data, but overall agreement for angular momentum and
spin. It remains to be seen if this offset is due to the different
renormalization procedures of the lattice operators or due to the
different $N_f$.


{\small
\bigskip
\begin{spacing}{1.0}
The numerical calculations have been performed on the APEmille, apeNEXT systems 
and PAX cluster at NIC / DESY (Zeuthen, Germany), the IBM BlueGene/L at EPCC
(Edinburgh, UK), the BlueGene/P (JuGene) and the Nehalem Cluster (JuRoPa) at NIC
(J\"ulich, Germany), and the SGI Altix and ICE 8200 systems at LRZ (Munich,
Germany) and HLRN (Berlin-Hannover, Germany). We have made use of the Chroma
software suite \cite{Edwards:2004sx}. The BlueGene codes were optimised with
Bagel \cite{Bagel}. This work has been supported in part by the DFG (SFB/TR 55,
Hadron Physics from Lattice QCD) and the EU under grants 238353 (ITN STRONGnet)
and 227431 (HadronPhysics2). A.St acknowledges support by the European
Reintegration Grant (FP7-PEOPLE-2009-RG, No.256594). JZ is supported by the
University of Adelaide and the Australian Research Council through a Future
Fellowship (FT100100005).
\end{spacing}
}
\vspace{-0.2cm}


\providecommand{\href}[2]{#2}\begingroup\raggedright\endgroup

\end{document}